\documentclass[11pt]{article}

\usepackage{subfloat}
\usepackage[utf8]{inputenc}
\usepackage[english]{babel}
\usepackage{enumerate}
\usepackage[shortlabels]{enumitem}
\usepackage{amsmath}
\usepackage{caption}
\usepackage{graphicx}
\usepackage{subcaption}
\usepackage{nicefrac}
\usepackage{bbm}
\usepackage[sorting=none]{biblatex}
\usepackage{multirow}
\numberwithin{equation}{section}
\title{Machine learning in/for blockchain: Future and challenges}
\author{Fang Chen\footnote{Ph.D. student, Department of Industrial Engineering, Purdue University.}, Hong Wan\footnote{Associate Professor, Department of Industrial and System Engineering, North Carolina State University.}, Hua Cai\footnote{Assistant Professor, Department of Industrial Engineering, Purdue University.}, and Guang Cheng\footnote{Corresponding Author. Professor, Department of Statistics, Purdue University. Guang Cheng gratefully acknowledges NSF DMS-1712907, DMS-1811812, DMS-1821183,and Office of Naval Research, (ONR N00014-18-2759).}}
\date{}

\begin{document}
\maketitle
\begin{center}
    \textbf{Abstract}
\end{center}
Machine learning and blockchain are two of the most noticeable technologies in recent years. The first one is the foundation of artificial intelligence and big data, and the second one has significantly disrupted the financial industry. Both technologies are data-driven, and thus there are rapidly growing interests in integrating them for more secure and efficient data sharing and analysis. In this paper, we review the research on combining blockchain and machine learning technologies and demonstrate that they can collaborate efficiently and effectively. In the end, we point out some future directions and expect more researches on deeper integration of the two promising technologies. 


\section{INTRODUCTION}
A blockchain is a shared, distributed public ledger that stores transaction data in a chain of sequential blocks (Dinh \& Thai, 2018). The data (block) are time-stamped and validated before adding to the chain. Each block contains information from the previous one. The mathematical structure for storing data makes it nearly impossible to fake (MIT Technology Review Editors, 2018). Thanks to the legacy of cryptocurrency, the term "blockchain" has transformed from a cryptography terminology to a buzz word. Many people believe that cryptocurrency IS blockchain. This is incorrect. While blockchain is the foundation of cryptocurrency, the applications of the blockchain technology are much wider. Scenarios involving data validating, auditing, and sharing can all consider applying blockchains.    

In this paper, we review the research on combining blockchain and machine learning technologies and demonstrate that they can collaborate efficiently and effectively. Machine learning is a general terminology that includes variety of methods, machine learning, deep learning and reinforcement learning. These methods are the core technology for big data analysis (Buhlmann et al., 2019). As a distributed and append-only ledger system, the blockchain is a natural tool for sharing and handling big data from various sources through the incorporation of smart contracts (i.e., a piece of code that will execute automatically in certain conditions). More specifically, blockchain can preserve data security and encourage data sharing when training and testing machine learning models. Also, it allows us to utilize distributed computing powers (for example, IOT), for developing on-time prediction models with various sources of data. This is especially important for deep learning procedures which require a  tremendous amount of computational power. On the other hand, blockchain systems will generate a huge amount of data from different sources, and the distributed systems are harder to monitor and control than the centralized ones. Efficient data analysis and forecasting of the system behaviors are critical for optimal blockchain mechanism designs. In addition, machine learning can facilitate the data verification process and identifying malicious attacks and dishonest transactions in the blockchain. The interdisciplinary research on combining the two technologies is of great potential. 

In this paper, we review articles that are either using machine learning techniques to study the blockchain system/structure itself or implementing blockchain techniques to improve machine learning, e.g., collaborative/distributed learning. The reviewed papers are summarized in Table 1 below. For papers that apply machine learning and blockchain techniques separately to various areas, we do not include them in our review but list some of them in Table 2 below. In the rest of this paper, we first review basic structure and terminology of blockchain in Section 2. The review is by no means exhaustive, but sufficient for Sections 3, 4, and 5 that introduce how different machine learning methods can be incorporated into the blockchain system. Our work is concluded by Section 6 to discuss potential research directions and challenges that are arose from ongoing and future fusion of machine learning and blockchain.

\begin{table}[]
\resizebox{\textwidth}{!}{%
\begin{tabular}{cll}
\hline
Method                                                                                          & \multicolumn{1}{c}{Application}   & \multicolumn{1}{c}{Paper}                                                                                                                             \\ \hline
\begin{tabular}[c]{@{}c@{}}Supervised/Unsupervised Learning\\ without Deep Methods\end{tabular} & Transaction Entity Classification & \begin{tabular}[c]{@{}l@{}}Yin \& Vatrapu (2017),  Jourdan et al. (2018), \\ Akcora et al. (2020)\end{tabular}                                        \\ \cline{2-3} 
                                                                                                & Bitcoin Price Prediction          & \begin{tabular}[c]{@{}l@{}}Jourdan et al. (2018), Shah \& Zhang (2014), \\ Akcora et al. (2019),  Abay et al.  (2019), Dey et al. (2020)\end{tabular} \\ \hline
\begin{tabular}[c]{@{}c@{}}Supervised Learning \\ with Deep Methods\end{tabular}                & Privacy and Security Preserving   & \begin{tabular}[c]{@{}l@{}}Harris \& Waggoner (2019), Chen et al. (2018), \\ Zhu, Li, \& Yu (2019)\end{tabular}                                       \\ \cline{2-3} 
                                                                                                & Computation Power Allocation      & Luong et al. (2018)                                                                                                                                   \\ \cline{2-3} 
                                                                                                & Cryptocurrency Price Prediction   & \begin{tabular}[c]{@{}l@{}}Mcnally, Roche, \& Caton (2018), Lahmiri \& Bekiros (2019),\\ Alessandretti et al. (2018)\end{tabular}                     \\ \hline
Reinforcement Learning                                                                          & IoT Network                       & Liu, Lin, \& Wen (2018)                                                                                                                               \\ \cline{2-3} 
                                                                                                & Bitcoin Mining                    & \begin{tabular}[c]{@{}l@{}}Eyal \& Sirer (2014), Sapirshtein, Sompolinsky \& Zohar (2017),\\ Wang, Liew, \& Zhang (2019)\end{tabular}                 \\ \hline
\end{tabular}%
}
\caption{Summary of Papers in the Review}
\label{tab:1}
\end{table}

\begin{table}[]
\resizebox{\textwidth}{!}{%
\begin{tabular}{ll}
\hline
Area        & Exemplary Sources                                                                                                                                                                                                                                                                                            \\ \hline
Healthcare  & \begin{tabular}[c]{@{}l@{}}Mamoshina et al. (2017), Juneja \& Marefat (2018), Okalp et al. (2018), \\ Zheng et al. (2018), Firdaus et al. (2018), Wang et al. (2018), \\ Vyas, Gupta, \& Yadav (2019), Bhattacharya et al. (2019), \\ Agbo, Mahmoud, \& Eklund (2019), Khezr et al. (2019).\end{tabular}     \\ \hline
IoT Related & \begin{tabular}[c]{@{}l@{}}Liu, Lin, \& Wen (2018), Xiong \& Xiong (2019), Lee \& Ryu (2018),  \\ Qin et al. (2019), Ozyilmaz, Dogan, \& Yurdakul (2018), Singla, Bose, \& Katariya (2018),  \\ Shen et al. (2019), Li et al. (2019), Rathore, Pan,  \& Park (2019), Ferrag \& Maglaras (2020).\end{tabular} \\ \hline
\end{tabular}%
}
\caption{Summary of Some Less Relevant Papers}
\label{tab:2}
\end{table}

\section{REVIEW ON BLOCKCHAIN}

A blockchain, literally speaking, is just a chain of digital blocks. Each block contains a certain amount of data; and the chain connects these data to form a distributed database. A newly created block includes multiple transactions collected from nodes and broadcasts to every node on the network. It can be accepted and added to the blockchain by nodes that have the same consensus protocol. Each added block includes information of the previous block in the chain. Hence, if the block is changed, all blocks {\em before} this block will be invalid as well. The strategies to reach agreement of the new block (consensus) vary in different types of blockchain. The mathematical structure of the blockchain implies two essential properties: (i) the data (in block) is immutable (MIT Technology Review Editors, 2018); (ii) the distributed network with consensus allows users to communicate directly with each other and download a copy of the current ledger, which means that there is continuous monitoring and redundancy of the data in the network. Therefore, the blockchain is more robust to individual outrages and attacks.

Depending on who can access to the blockchain and who can validate the data, the blockchain can be classified into public chains, private chains, and consortium chains (Zheng et al., 2018). The comparison of three different types of blockchains is shown in the Table 3. 

\begin{table}[]
\resizebox{\textwidth}{!}{%
\begin{tabular}{l l l l}
\hline
\textbf{Attribute}       & \textbf{Public}           & \textbf{Private}      & \textbf{Consortium} \\ \hline
Who run/manage the chain & All miners                & One organization/user & Selected users      \\ \hline
Permission to Access     & No                        & Yes                   & Yes                 \\ \hline
Security                 & Nearly impossible to fake & Could be tampered     & Could be tampered   \\ \hline
Efficiency               & Low                       & High                  & High                \\ \hline
Centralized              & No                        & Yes                   & Partial             \\ \hline
Example                  & Bitcoin, Ethereum         & IBM HyperLedger       & Quorum              \\ \hline
\end{tabular}%
}
\caption{Comparison of three types of blockchains}
\end{table}
In what follows, we use the bitcoin system, which is the most known blockchain application, as an example to demonstrate how blockchain works in detail. Typically, an end-to-end blockchain-based transaction needs to be validated at two different levels, the node level and the block level. The transaction is first verified between two nodes (Zheng et al., 2018). Then a unique digital signature, which is a hash wrapping all information of the transaction, is created. The digital signature that represents the transaction is submitted to the transaction pool and is waited to be added to a new block. Before the new block is accepted by the blockchain network, it is required to be validated by other miners on the network through the Proof-of-Work (PoW) consensus protocol. The PoW process includes aggregating a set of transactions to the new block and finding a hash value that is lower than the target value (Ghimire \& Selvaraj, 2018). The new block is only accepted by the network if transactions are valid and unspent. Other nodes continue working on creating the new block using the hash from the previous block (Nakamoto, 2008). 

Since the probability that finding a new valid block is extremely low and the PoW process requires a huge amount of computing power and a high consumption of the electricity, miners tend to collaborate with each other and form a mining pool. After participating in a mining pool, individual miner could receive a steady reward and significantly lower the risk. On the other hand, the mining pools usually charge membership fees to each participant and allocate rewards to each miner according to their own rewards sharing mechanisms (Bhaskar \& Lee, 2015). Some common reward allocation mechanisms in practice are Pay-Per-Last-N-Shares (PPLNS) (Qin, Yuan, \& Wang, 2019) and Full-Pay-Per-Share (FPPS) (Zhu et al., 2018). Another popular public chain is Ethereum (Wood, 2014), which allows users to send not only digital coins but also smart contracts (Wohrer \& Zdun, 2018). In order to reduce the energy consumption on the validation, Ethereum plans to switch its consensus protocol from the PoW to the Proof-of-Stake (PoS) gradually (Saleh, 2020).

\section{SUPERVISED/UNSUPERVISED LEARNING WITHOUT DEEP METHODS}\label{sec:4}
In this section, we review several applications of machine learning for the blockchain. Specifically, Section~\ref{sec:4.1} reviews three studies regarding transaction entities classification (Yin \& Vatrapu, 2017; Jourdan et al., 2018; Akcora et al., 2020) with different purposes. One focuses on the recognition of cybercriminal entities using supervised learning (Yin \& Vatrapu, 2017) as well as topological data analytics (TDA) method (Akcora et al., 2020), while another on the recognition of common categories of entities for most transactions (Jourdan et al., 2018). Section~\ref{sec:4.2} reviews Bitcoin price prediction from different perspectives such as probabilistic graphic models (Jourdan et al., 2018), Bayesian regression (Shah \& Zhang, 2014) and feature selection on the blockchain topological structure using Granger causality and TDA (Akcora et al., 2019; Abay et al., 2019; Dey et al., 2020).

\subsection{Transaction entity classification}\label{sec:4.1}

In the Bitcoin network, it is crucial to recognize entities behind those potentially illegal ones. The study of identifying entities behind addresses is called address clustering (Harrigan \& Fretter, 2016). Yin \& Vatrapu (2017) apply supervised learning to classify entities of transactions that may involve in cybercriminal activities. The classification model is trained based on 854 observations with categorical identifiers and then applied to study 10000 uncategorized observations that take 31.62\% of unique addresses and 28.99\% of total coins in the overall Bitcoin blockchain. The categorical identifiers represent 12 classes of entities, five of which are related to cybercriminal activities. Thirteen classifiers from the Python machine learning package ``scikit-learn'' are applied. By comparing accuracy scores of all classifiers, it is found that Random Forests (77.38\%), Extremely Randomised Forests(76.47\%), Bagging (78.46\%) and Gradient Boosting (80.76\%) stand out as the best four classifiers. After further comparing precision, recall, and f1 score of these classifiers, bagging and gradient boosting stand out, which are then applied to analyze the 10000 observations. The classification outcome shows that 5.79\% (3.16\%) addresses and 10.02\% (1.45\%) coins are from cybercriminal entities according to the bagging method (gradient boosting method).

Bitcoins are found to be a common way to make the ransomware payment. In order to detect addresses related to ransomware payment, Akcora et al. (2020) apply a topological data analysis (TDA) approach to generate the bitcoin address graph by first grouping similar addresses into nodes and then putting common addresses between two nodes into the set of edges. The TDA is an approach commonly used for dimension reduction. It represents the data set in a graph by first dividing data to sub-samples based on different filtration criteria and then clustering similar points in each sub-sample. The Bitcoin transaction graph model is a directed graph, denoted as $\mbox{G} = (V, E, B)$, where $V$ is the set of vertices, $E$ is a set of edges and $B = \{\mbox{Address}, \mbox{Transaction}\}$ is a set of node types. By using six graph features extracted for each address, a TDA Mapper method is applied to create six filtered cluster tree graphs. After calculating the number of ransomware addresses in each cluster, denoted as $V$, a suspicion score is assigned to a new address. The suspicion scores of addresses in the cluster are set to be 0 initially. It increments by one if inclusion and size thresholds are satisfied as follow: (1) the inclusion threshold, denoted as $\epsilon_1$, times the total amount of labeled ransomware addresses is less than $V$; (2) the size threshold, denoted as $\epsilon_2$, times the number of labeled ransomware addresses in the cluster is greater than the number of all addresses in the cluster. Suspicious addresses are then filtered by a quantile threshold, denoted as $q$, when their suspicious scores are higher than the quantile threshold. The result indicates that the best TDA model with $\epsilon_1=0.05, \epsilon_2 = 0.35, q = 0.7$ outperforms random forest (RF), and XGBoost in new ransomware addresses prediction.

Jourdan et al. (2018) are interested in classifying entities of transactions into four most common categories: Exchange, Service, Gambling, Mining Pool, based on data collected from 97 sources (Ermilov, Panov, \& Yanovih, 2017). The goal of classification is to assist in selecting an appropriate prediction model that is built according to categories of transactions (Jourdan et al., 2018). The applied classification method is a gradient boosted decision tree algorithm along with a Gaussian process based optimization procedure that determines optimal hyperparameters. Table 4 concludes that accuracy's in Exchange, Gambling, and Service categories are high. However, the accuracy in the Mining Pool category is poor. This may indicate that mining activities may not be appropriate as an independent label.

\begin{table}[]
\centering
\begin{tabular}{llll}
\hline
Category & Accuracy & $F_1$ & Precision \\ \hline
Exchange & 0.94     & 0.92  & 0.91      \\
Gambling & 0.95     & 0.97  & 1.00      \\
Mining   & 0.50     & 0.67  & 1.00      \\
Service  & 0.95     & 0.88  & 0.83      \\
Overall  & 0.92     & 0.91  & 0.92      \\ \hline
\end{tabular}
\caption{Classification Performance (Jourdan et al., 2018)}
\label{tab:4}
\end{table}

\subsection{Bitcoin price prediction}\label{sec:4.2}

UTXOs record the number of Bitcoins in transactions, which enables us to track buying and selling information to predict the Bitcoin price. Another contribution of Jourdan et al. (2018) is to forecast the value of UTXOs by creating probabilistic graphical models. The first model is called the Block-transaction address model (BT-A) that is a stationary graphic model of a Bitcoin block with conditional dependency structures. As an extension of BT-A, a Block-transaction entity-address model (BT-EA) is further developed by adding a categorical entity to each address. In terms of MSE, RMSE, MAE, RMAE, simulation results in Table 5 show that this extension significantly outperforms the BT-A model in all categories except for Exchange.

\begin{table}[]
\centering
\begin{tabular}{llllll}
\hline
Metric & BE-TA & BE-TA & BE-TA & BE-TA & BT-A \\ \cline{2-6} 
       & E     & S     & G     & M     & All  \\ \cline{2-6} 
MSE    & 1.22  & -0.3  & -0.02 & 0.06  & 1.12 \\
RMSE   & 125   & 53.3  & 1.15  & 5.19  & 90.5 \\
MAE    & 15.6  & 0.94  & 0.20  & 2.42  & 7.47 \\
RMAE   & 1.82  & 1.74  & 1.86  & 1.93  & 1.69 \\
NRMSE  & 1.34  & 1.28  & 1.42  & 1.22  & 1.29 \\ \hline
\end{tabular}
\caption{BT-A and BT-EA Performance (Jourdan et al., 2018)}
\label{tab:5}
\end{table}

The dependence structure of the BT-A model to obtain the output UTXOs values, denoted as $V_{o,u}$, is illustrated in Figure 1. Here is some explanation. The BT-A model starts with computing the number of available UTXOs for the $i^{th}$ input address $A_i$, denoted as $k_{A_i}^{UTXO}$. For each input address, the number of UTXOs used in a transaction is uniformly drawn from 1 to $k_{A_i}^{UTXO}$ with the corresponding UTXO value, denoted as $V_{i,u}$. The total input value of a transaction is calculated by summing the input UTXOs value of each input address, denoted as $V_t = \sum V_{i,u}$, and the value of an output UTXO is uniformly drawn from 1 to total transaction value minus validation fee.

\begin{figure}
    \centering
    \includegraphics[scale=0.7]{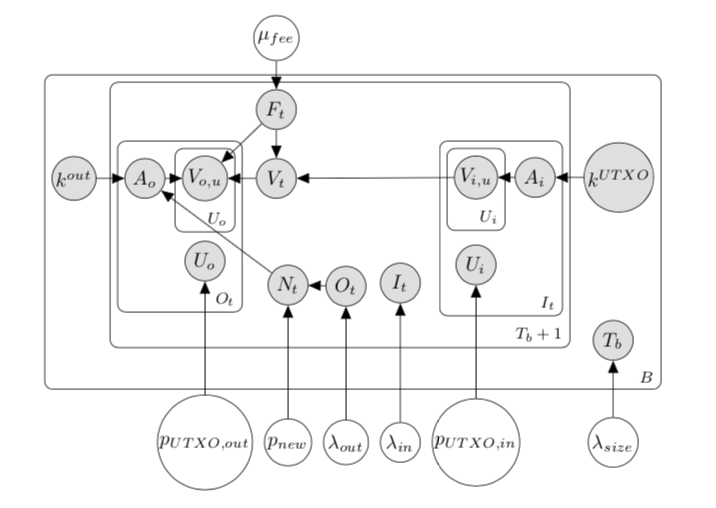}
    \caption{Block-transaction Address Model (Jourdan et al., 2018)}
\end{figure}

To predict the Bitcoin price, Shah and Zhang (2014) apply the Bayesian regression for the ``latent source model'' that is a nonparametric model for time series binary classification. The latent source model framework is described in the Chen, Nikolov, \& Shah (2013) and Bresler, Chen, \& Shah (2014), which latent sources are time series with binary labels. Specifically, the model is described as $K$ distinct unknown latent sources, $s_1, ..., s_K$ generated from a latent distribution over  $\{1, ..., K\}$ with probabilities $\{\mu_1,..., \mu_k\}$; $K$ latent distributions, denoted as $P_1, ..., P_K$. Labeled data are generated from sample index $T \in \{1, ..., K\}$ with $P(T = k) = \mu_k$. The model that predicts $y$ given $x$ refers to the equation below.
\begin{equation}\label{eq:12}
\begin{split}
P(y|x) & = \sum_{k=1}^T P(y|x, T=k) P(T=k|x) \\
    & = \sum_{k=1}^T P_k(y) \mbox{exp}\left(-\frac{1}{2}||x-s_k||_2^2\right)\mu_k
\end{split}
\end{equation}
where $T \in \{1, ..., K\}$ is the sample index with $P(T = k) = \mu_k$
\newline Due to lacks of information of latent parameters, empirical data is used as proxy for estimating $P(y|x)$. The expectation of $P(y|x)$ can be estimated as follows:

\begin{equation}\label{eq:13}
E[y|x] = \frac{\sum_{i=1}^n y_i \exp(-\frac{1}{4}||x - x_i||_2^2)}{\sum_{i=1}^n \exp(-\frac{1}{4}||x - x_i||_2^2)}
\end{equation}
The future average price change is determined by price changes over three periods of historical data: previous 30 minutes sample, 60 minutes sample and 120 minutes sample, denoted as $\Delta p^j, j = 1, 2, 3$. Each $\Delta p^j$ is calculated by~(\ref{eq:13}). Then $\Delta p$ over a 10-second period is formulated as
\begin{equation}\label{eq:14}
\Delta p = w_0 + \sum_{j=1}^3 w_j\Delta p^j + w_4r
\end{equation}
\begin{itemize}
    \item $w_0, w_1, w_2, w_3, w_4$ are weights to be estimated.
    \item $r = (v_b-v_a) / (v_b+v_a)$, where $v_b, v_a $ are the top 60 orders of total buying and selling volume.
\end{itemize}
We would like to point out that in order to apply formula~(\ref{eq:14}), it is crucial to verify the stationarity of the price data, which was unfortunately not done in the referenced paper. The trading strategy for each user is designed as ``buy one bitcoin when $\Delta p> t$; sell one bitcoin when $\Delta p< -t$; otherwise holding the current number of bitcoin when $-t\le \Delta p \le t$.'' Here, $t$ is a pre-specified threshold. The designed prediction model is trained by data gathered from Okcoin before May 2014 and is tested by data after that. It is found that increasing $t$ leads to an increase of the average profit per trade.

Besides using the Bayesian regression to predict the bitcoin price, the selection of input features is also important to the performance of the prediction. To better characterize input features, Akcora et al. (2019) introduce a concept of graphic chainlet, which describes the local topological features of Bitcoin blockchain, to explore impacts of the Bitcoin blockchain structure on Bitcoin price formation and dynamics. A transaction-address graph representation of the Bitcoin blockchain is shown in Figure 2. Circle vertices represent input and output addresses. A square vertex indicates the transactions and edges stand for UTXOs (a transfer of Bitcoins). A chainlet model represents $x$ input UTXOs and $y$ output UTXOs involving in a transaction, denoted as $C _{x \rightarrow y}$. All chainlet and chainlet clusters clustered by various criteria are evaluated by the Granger causality test (Granger, 1969). The result concludes that the split chainlet cluster defined as when $y < x < 20$, individual chainlet (e.g., $C_{1 \rightarrow 7}$, $C_{6 \rightarrow 1}$, $C_{3 \rightarrow 3}$), extreme chainlets (e.g., $C_{20 \rightarrow 2, 3, 12, 17}$), clusters according to Cosine Similarity (e.g., $C_{9 \rightarrow 11}$, $C_{3 \rightarrow 17}$, $C_{8 \rightarrow 14}$, $C_{1 \rightarrow 1}$) are significant to the Bitcoin price formation and dynamics. A price prediction model is further developed using significant chainlets. 

\begin{figure}
    \centering
    \includegraphics[scale=0.4]{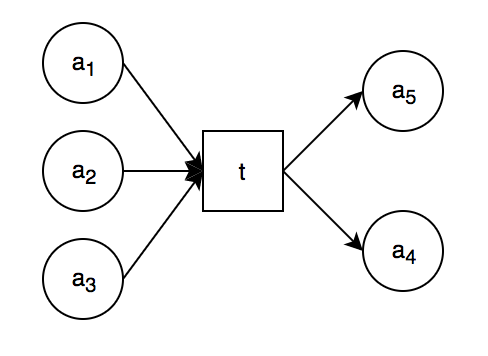}
    \caption{A Transaction-Address Graph}
\end{figure}

Chainlet model studies topological features from a single transaction aspect and only takes the number of input and output UTXOs into account. Abay et al. (2019) extend the chainlet model to a new graphic model ``ChainNet'' that further considers topological features from the aspects of the number of distinct chainlets and the amount of coins transferred by chainlets. More specifically, from the perspective of all transactions, an occurrence matrix is created to count the number of distinct chainlets among all transactions. An amount matrix records the sum of Bitcoins transferred for distinct chainlets. By considering both occurrence and the amount bitcoins transferred in a transaction, an occurrence matrix with a threshold, denoted as $O^{\epsilon}, \epsilon \in \{0, 10, 20, 30, 40, 50\}$, is created to count the number of distinct $C_{i \rightarrow j}$ that is larger than $\epsilon$. Different thresholds result in different values of $O^{\epsilon}$, which are considered as Filtration Features (FL) input in the prediction model. Betti sequences and Betti derivatives for the blockchain network are also considered as features in the model. A sliding prediction approach associated with parameters of prediction horizon, window length and training length is applied to train the time series prediction model. According to simulation results, ChainNet adopts Betti model features and FL features for short and long term prediction, respectively, to obtain a better performance.

Besides considering the effects of features of the Bitcoin topological structure on Bitcoin price formation and dynamics, topological features of other types of cryptocurrencies may also affect the Bitcoin price. Dey et al. (2020) evaluate the Bitcoin price formation and dynamics using the Chinalet model according to the topological features on the joint of Bitcoin and Litecoin. Specifically, the occurrences of distinct chainlets in Bitcoin and Litecoin, denoted as $O^B_{x \rightarrow y}$ and $O^L_{x \rightarrow y}$ respectively, are considered. The amounts of coins transferred in Bitcoin and Litecoin, denoted as $A^B_{x \rightarrow y}$ and $A^L_{x \rightarrow y}$ respectively, are also included. Granger causality tests (Granger, 1969) with 1 to 5 lag effects are applied to assess the significance of chainlets. The result concludes that the occurrence of chainlets in the Litecoin ($O^L_{3 \rightarrow 3}$, $O^L_{4 \rightarrow 4}$, $O^L_{4 \rightarrow 5}$, $O^L_{3 \rightarrow 6}$) is significant to the price for all five lag effects in the Granger causality. Also, occurrence and amount of chainlets in the Bitcoin ($O^B_{20 \rightarrow 2, 3, 12}$, $O^B_{1 \rightarrow 7}$, $A^B_{20 \rightarrow 12, 20}$, $A^B_{3 \rightarrow 4}$) are also important to the price for all five lag effects. 

Although there are other studies related to Bitcoin price prediction using machine learning methods, e.g., Greaves \& Au (2015), Jiang \& Liang (2017), Jang \& Lee (2018), and Sun, Liu, \& Sima (2018), it is hard to include all papers in the review. As a result, we will move on to review more articles in prediction of cryptocurrency price using deep learning in Section~\ref{sec:5}.

\section{SUPERVISED LEARNING WITH DEEP METHODS}\label{sec:5}

In this section, we turn to the application of deep learning. In Section~\ref{sec:5.1}, three privacy-preserving collaborative learning frameworks (Harris \& Waggoner 2019; Chen et al., 2018; Zhu, Li, \& Yu, 2019) are reviewed. In Section~\ref{sec:5.2}, we review a deep learning work (Luong et al., 2018) that allocates computation resource to assist mobile blockchain mining. In Section~\ref{sec:5.3}, we focus on cryptocurrency price prediction (Mcnally, Roche, \& Caton, 2018; Lahmiri \& Bekiros, 2019) and digital portfolio management using Recurrent Neural Network (RNN) and Long-Short Term Memory (LSTM) models (Alessandretti et al., 2018). 

\subsection{Decentralized privacy-preserving collaborative learning}\label{sec:5.1}

Harris \& Waggoner (2019) build a decentralized collaborative learning framework with blockchain. The new designed framework extended by the previous two frameworks (Abernethy \& Frongillo, 2011;  Waggoner, Frongillo, \& Abernethy, 2015) is designed to collaboratively build a dataset and train a predictive model. The framework starts with letting the provider define a loss function and upload 10 out of 100 partial dataset with corresponding hashes. By using the smart contract that initially contains a model, other participants add their own data or uploading an update along with a deposit of 1 unit of currency until the end condition set by the provider is met. The provider uploads the rest of 90 partial datasets to evaluate participants' models. The better model tends to receive more rewards in the end.  

Chen et al. (2018) propose a framework called ``Learning Chain'' to preserve user's privacy by applying a decentralized version of the Stochastic Gradient Descent (SGD) algorithm and a differential privacy mechanism. The proposed framework contains three phases: blockchain initialization; local gradient computation; global gradient aggregation. In the first phase, a peer-to-peer network is set up with computing nodes and data holders. The second phase involves each data holder $P_k$ retrieving the current model from the block $t$, denoted as $w_t$, and computing its own local gradient. A differential privacy mechanism is then applied to generate a hidden local gradient, denoted as $\nabla g_k(w_t)^*$, by adding a noise factor to the local gradient. The message broadcasts a pseudo-identity of $P_k$, normalized hidden local gradient, denoted as $\nabla \widehat{g_k}(w_t)^*$, together with the norm of its un-normalized version to computing nodes on the network. In the final phase, after solving Proof-of-Work (PoW), the winner node selects top l-nearest local normalized gradients according to the cosine distance between each normalized local gradient and the sum vector of $\nabla g_k(w_t)^*$ to update the global gradient. The predictive model is updated by $w_{t+1}=w_t+\eta \nabla J(w_t)$, where $\nabla J(w_t)$ is the updated global gradient. 

``Learning Chain'' is trained and tested in three different data sets: synthetic data set; Wisconsin breast cancer data set; MNIST data set; using the Ethereum blockchain framework. There exists a trade-off between privacy and accuracy in the sense that decreasing the privacy budget leads to an increase of test errors on all data sets. This proposed model is further compared with the ``Learning ChainEX'', which is implemented with higher differential privacy and has similar test error. 

Zhu, Li, \& Yu (2019) develop a blockchain-based privacy-preserving framework to secure the share of updates in federated learning. The Federated Learning algorithm is developed by McMahan et al., (2017), which allows each mobile device to compute and upload updates to the global predictive model based on their local data sets. A security issue arises when there exist Byzantine devices in the network. In this case, the blockchain transaction mechanism is adopted to ensure the security of sharing and updating changes. Specifically, model updates are written in a blockchain transaction by nodes. Along with the digital signature of a node, a transaction broadcasts to other nodes information, including changes of hyperparameters and weights, public keys (participants' addresses). Other nodes validate the transaction and test updates according to their local data sets. If most nodes confirm that the performance score of the updated model is higher than the existing model under their local data sets, the updates are implemented into the current model.  

\subsection{Computing power allocation}\label{sec:5.2}

Luong et al. (2018) develop a deep learning-based auction algorithm for edge computing resources allocation to support mobile mining activities. The designed framework enables mobile device miners to submit their bid valuation profiles to one Edge Computing Service Provider (ECSP) for buying additional computing power. The valuation profile for miner $i$, denoted as $v_i$, is drawn from a distribution that assigns a higher value $v_i$ when its block size divided by initial computing capacity is larger. The ECSP evaluates all valuation profiles and maximizes its revenue in the following steps. 

An allocation rule is applied to map transformed valuation profiles, denoted as $\overline{v}_i:=\phi_i(v_i)$, to assignment probabilities using a Softmax function. The winner miner $i$ will pay the price $p_i := \phi_i^{-1}(\mbox{ReLU}(\max_{i \ne j}\overline{v}_j))$. In the end, the loss function of ECSP is defined as 
\begin{equation}\label{eq:29}
\widehat{R} (\mathbf{w},\beta)=-\sum_{i=1}^N g_i^{(\mathbf{w},\mathbf{\beta})}(\mathbf{v}^s)p_i^{(\mathbf{w},\mathbf{\beta)}}(\mathbf{v}^s) 
\end{equation}
where stochastic gradient descent (SGD) is applied. Here, $g_i$ is the assignment probability and $N$ is the number of miners. The above designed deep learning (DL) based auction mechanism is empirically compared to a regular auction mechanism. It is found that DL-based auction achieves higher revenue and converges to the optimal value faster than other mechanisms.

\subsection{Cryptocurrency price prediction}\label{sec:5.3}

For forecasting Bitcoin price, Mcnally, Roche, \& Caton (2018) compare performances of two deep learning algorithms, i.e., Recurrent Neural Network (RNN) and Long-Short Term Memory (LSTM). It is interesting to note that two hidden layers with 20 nodes per layer are sufficient in both models. Specifically, the RNN model adopts the tanh fcuntion as its activation function while LSTM applies tanh and sigmoid functions for different gates, which result in longer training time. The data set used to train and test LSTM and RNN models is the bitcoin price from Aug 19th, 2013 to July 19th, 2016. Features including the opening price, daily high, daily low, the closing price, hash rate, and mining difficulty are used in the model. The importance of features is evaluated by the Boruta algorithm, which is a wrapper built around the random forest classification algorithm. The traditional time series model, AutoRegression Integrated Moving Average (ARIMA), is empirically compared with these deep learning models. The simulation results show that LSTM, RNN, and ARIMA have similar accuracy, which are 52.78\%, 50.25\%, and 50.05\%. However, deep learning models have much lower RMSE values. In addition, the LSTM model is capable of recognizing long-term dependencies in contrast to the RNN model. 

In contrast with other studies mainly for predictive models, Lahmiri \& Bekiros (2019) instead conduct a chaotic time series analysis before building deep learning models. Hence, their first step is to calculate the largest Lyapunov exponent (LLE) and then apply detrended fluctuation analysis (DFA) to detect chaos characteristics of cryptocurrency price data without having the assumption of stationarity. Then a deep neural network (DLNN) model with LSTM implementation (Hochreiter \& Schmidhuber, 1997) and a generalized regression neural network (GRNN) model (Specht, 1991) are built to predict three types of cryptocurrency: Bitcoin, Digital Cash, and Ripple price. The number of data samples obtained for the model is 3006 Bitcoin, 1704 Digital Cash ,and 1357 Ripples. The authors create a many-to-many sequence prediction, which utilizes the first 90\% observations for training and the last 10\% observations for testing and out-of-sample forecasting. According to Figure 3 whose x-axis represents the time horizon and the y-axis represents the price, positive Hurst exponent (HE) value indicates long-memory features of data, and negative LLE value indicates training data is chaos. As a result, a short-term prediction model would be suitable for data. The simulation results claim that the LSTM model outperforms the GRNN model in all three cryptocurrencies' price predictions. Although the RMSE of the LSTM model is still high, the model demonstrates a similar trend to real price changes for all three cryptocurrencies.  

\begin{figure}
    \centering
    \includegraphics[scale=0.57]{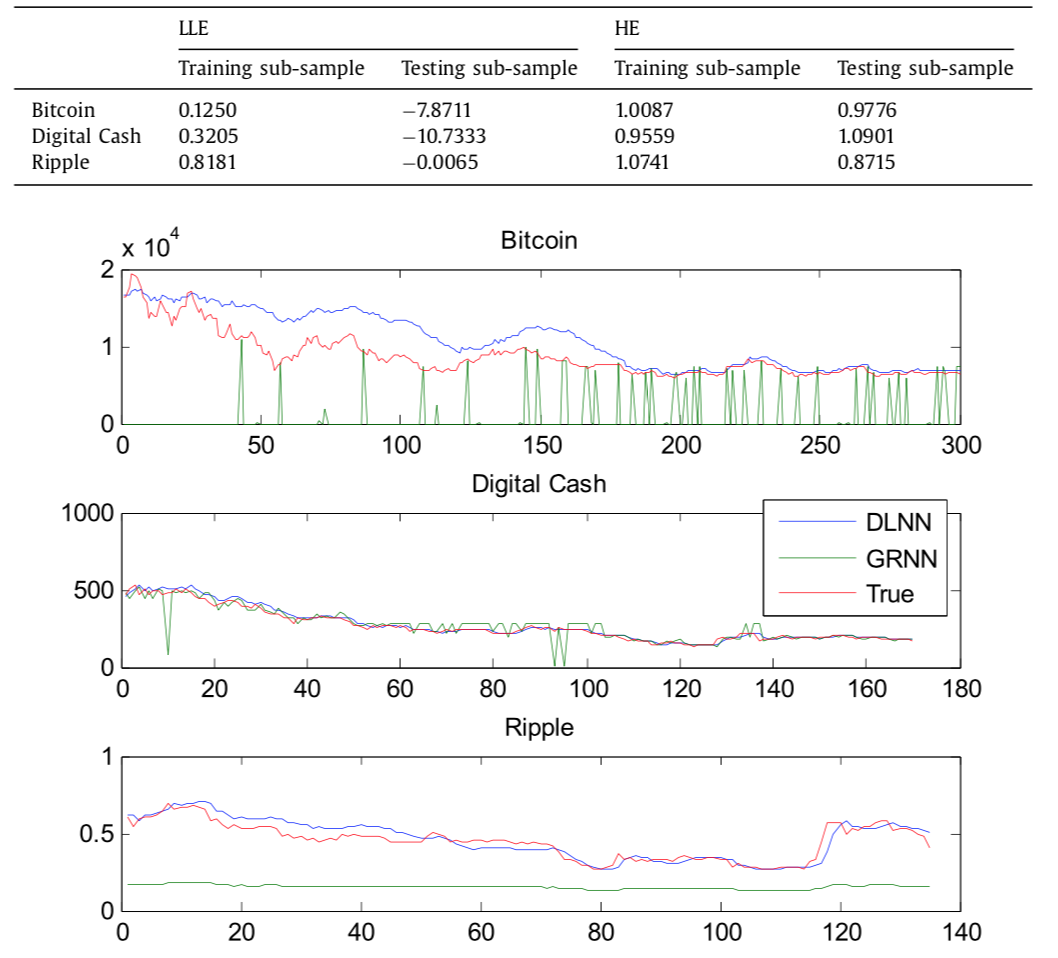}
    \caption{Chaotic Analysis and Prediction Result (Lahmiri \& Bekiros, 2019)}
\end{figure}

Besides cryptocurrency price prediction, Alessandertti et al. (2018) explore a portfolio analysis by forecasting daily prices of 1681 types of cryptocurrencies. Three models are developed to predict the prices of every kind of cryptocurrency. For each type $c$, the target is the return of investment (ROI) at each time $t_i \in \{0, ..., 895\}$, which is expressed as:
\begin{equation}\label{eq:30}
    \mbox{ROI}(c, t_i) = \frac{\mbox{price}(c,t_i)-\mbox{price}(c, t_i - 1)}{\mbox{price}(c, t_i-1)}
\end{equation}
Features considered are price, market capitalization, market share, rank, and volume. The first model is an ensemble of regression trees using XGboost, which features of each type of cryptocurrency are paring with prices of each type of cryptocurrency. The second model is a regression model by considering features of all kinds of cryptocurrency as a whole paired with prices of each type of cryptocurrency. The third model adopts RNN with LSTM implementation with the second model's features and target paring strategy. All models are one-step ahead forecasting. A portfolio is constructed based on the predicted prices. Model hyperparameters are optimized by maximizing either sharp ratio or geometric mean of the total return. The result concludes that all three models generate profits, and the optimization of parameters using the sharp ratio metric achieves a higher return. Another conclusion is that the first two models implementing gradient boosting with decision trees have higher accuracy for the short-term (5-10 days), while the third model adopting LSTM has a better prediction performance in the long term (around 50 days). 

\section{REINFORCEMENT LEARNING}\label{sec:6}

In this section, we first review a framework that incorporates reinforcement learning into blockchain to ensure the security of data collection, storage and processing in the IoT network (Liu, Lin, \& Wen, 2018). Secondly, we review two types of frameworks that study the Bitcoin blockchain mining activities. The first explores the potential of Bitcoin mining through the mobile network (Nguyen et al., 2020). While the second formulate a Markov decision process (MDP) for the blockchain mining activity (Eyal \& Sirer, 2014; Sapirshtein, Sompolinsky, \& Zohar 2017). The last work in the review applies a new reinforcement learning algorithm to find the optimal mining strategy (Wang, Liew, \& Zhang, 2019). 

\subsection{IoT}
Liu, Lin, \& Wen (2018) propose a framework to secure data collection and sharing among mobile terminals (MTs) on the IoT network. The framework consists of two phases: data collection and data sharing. In the first phase, each MT, denoted as $m$, adopts multi-agent deep reinforcement learning (DRL) to maximize efficacy of data collection. The state space is defined as $S = \{S_1, S_2, S_3\}$. Here, $S_1 = \{(x^k, y^k), (x^c, y^c)\}$ is a set of state which represents coordinates of $k$ Point-of-Interest (PoIs) and $c$ obstacles in the environment, denoted as $E_x \times E_y$, where $x \in [0, E_x], y\in [0, E_y]$; $S_2$ stands for MTs' coordinates and $S_3$ represents sensing time $h_t(k) \in [0, t]$ for the $i$-th POI. Action space consists of moving direction, denoted as $\theta_t^m$, and moving distance, denoted as $l_t^m$. Thus, it is written as $A = \{(\theta_t^m, l_t^m) \mid \theta_t^m \in [0, 2\pi), l_t^m \in [0, l_{max})\}$. The reward $r_t^m$ is given as
\begin{equation}\label{eq:6.1}
r_t^m=\frac{w_t b_t^m}{\alpha b_t^m+\kappa l_t^m}
\end{equation}
where $b_t^m$ is the amount of collected data, $\alpha$, $\kappa$ are the energy consumption per collected data and per travelled distance; $w_t$ is the achieved geographical fairness, calculated by $w_t = \frac{(\sum_{k=1}^K h_t(k))^2}{k\sum_{k=1}^K h_t(k)^2}$ Each MT is implemented by four deep neural networks and actor-critic algorithm is applied to maximize the reward. 

After MTs finish the data collection, they share data through an Ethereum blockchain network. However, the first step would be to send data to the certificate authority (CA) for verification. Once CA verifies the ownership of MTs' data and checks the consistence of received data and original data stored in the terminal, a digital signature is generated and sent back to the MT. As a result, the MT is able to broadcast its transaction request consisting of digital signature of CA, original data and its public key to other nodes on blockchain network to be further validated. By comparing to randomly moving MTs, MTs implemented DRL collect much more data but consume more energy. The blockchain-based data sharing framework can still store all data sent by MTs even under Dos attack. 

\subsection{Bitcoin mining}
As we discussed in earlier sections, the blockchain mining costs a huge amount of computing power, so it is nearly impossible to apply the blockchain to the mobile system. Nguyen et al. (2019) propose a mobile edge computing (MEC) based blockchain network to assist mobile users (MUs) offloading mining tasks to the MEC server. Specifically, the state space is defined as $s^t = \{D_1^t, D_0^t, g^t\}$, where $D_1^t, D_0^t$ are new and buffered transaction data at the time $t$ separately; $g^t$ is the power gain by miner $n$ offloads the task $m$ to the MEC server. The action space is expressed as $a^t = x_{nm}^t$, where $x_{nm}^t \in \{0, 1\}$ stands for the $n^{th}$ MU processes $m$ mining tasks locally or offloading $m$ tasks to the MEC server, respectively. The goal of a miner is to maximize the privacy level $P^t$ defined in He et al. (2017) and minimize the sum of cost of energy and time consumption. The system reward is formulated as $r^t(s,a) = P^t(s, a) - C^t(s, a)$, where $P(s, a)$ is the privacy level and $C(s, a)$ is the sum of the cost of the electricity and the computing power. A value-based method, Q-learning, and deep q learning are applied to update the Q value. The result concludes that although the convergence speed for Q-learning and deep q learning are almost the same, agents trained by the deep q learning model receive higher total rewards. 

Although mining bitcoins could generate a big revenue by selling bitcoins, the cost of mining is also high due to the high consumption of electricity. People now are interested in finding the optimal mining strategy to maximize their profits. Since Bitcoin mining is able to be modeled as a Markov decision process (MDP) that contains a enormous number of states, reinforcement learning is applied to study the MDP of bitcoin mining. The MDP for Bitcoin mining is first proposed by Eyal \& Sirer (2014) and then is extended by Sapirshtein, Sompolinsky, \& Zohar (2017). The environment assumes that the block generation time follows Poisson distribution and it is independently with each other. The new block is created by an honest agent with probability $(1 - \alpha)$, while the new block is obtained by the adversary agent, also known as an attacker, with probability $\alpha$. The adversary may hide some blocks on its own private chain, but the blockchain is always the longest public chain. The state space of the MDP is defined as $(a, h, fork)$, where $a$ represents the number of blocks on the adversary's private chain; $h$ represents the number of blocks on the public chain; fork is an environment variable that has three values, which are $(irrelevant, relevant, active)$. $(a, h, irrelevant)$ denotes the case when previous state is $(a-1, h)$ and $match$ action is feasible, i.e., the last mined block accepted by the chain is mined by the adversary miner; $(a, h, relevant)$ denotes the case when previous state is $(a, h-1)$ and $match$ action is infeasible, i.e., the last mined block is mined by the honest miner; $active$ refers to the case that the network is broken into two branches containing the same number of blocks. When $fork = active$, the probability that follows the honest block is $\gamma$ and the probability that follows the adversary block is $1 - \gamma$. The action space, defined as $A = (Adopt, Override, Match, Wait)$, contains four actions. The $Adopt$ refers to an agent always mines mines the last block on the public chain and do not have any blocks on its private chain. The $Override$ becomes feasible when the number of blocks on the private chain is more than the number of blocks on the public chain. In other words, all blocks on the private chain are published to replace the existing public chain. The $Match$ action refers to the adversary agent releases the same number of blocks as the current public chain, which creates a fork on the public chain. The $Wait$ action is always feasible, which the adversary agent keeps mining on its private chain and not releasing any new blocks to the public chain. The transition probability matrix is shown in Table 6. Since the honest agent is consider as a part of environment, we only focus on finding the optimal strategy for adversary agents. The number of blocks on the public chain is considered as rewards. The reward is formulated as two dimensions, which are the number of blocks mined by honest agents and adversary agents separately. The reward function then considers relative reward instead of a absolute reward. The objective function is defined as~\ref{eq:6.2}.

\begin{equation}\label{eq:6.2}
f(s,a)=\frac{q^{a}(s,a)}{q^{a}(s,a)+q^{h}(s,a)}
\end{equation}

\begin{table}[]
\centering
\resizebox{\textwidth}{!}{%
\begin{tabular}{llll}
\hline
State, Action              & State + 1                 & Transition Prob.                 & Reward     \\ \hline
$(a, h, ), adopt$          & $(1, 0, irrelevant)$      & $\alpha$                         & $(0 ,h)$   \\ \cline{2-4} 
                           & $(0, 1, irrelevant)$      & $1 - \alpha$                     & $(0 ,h)$   \\ \hline
$(a, h, ), override$       & $(a - h, 0, irrelevant)$  & $\alpha$                         & $(h+1 ,0)$ \\ \cline{2-4} 
                           & $(a - h -1, 1, relevant)$ & $1 - \alpha$                     & $(h+1 ,0)$ \\ \hline
$(a, h, irrelevant), wait$ & $(a + 1, h, irrelevant)$  & $\alpha$                         & $(0 ,0)$   \\ \cline{2-4} 
$(a, h, relevant), wait$   & $(a, h+1, relevant)$      & $1 - \alpha$                     & $(0 ,0)$   \\ \hline
$(a, h, active), wait$     & $(a + 1, h, active)$      & $\alpha$                         & $(0 ,0)$   \\ \cline{2-4} 
$(a, h, relevant), match$  & $(a - h, 1, relevant)$    & $\gamma \dot (1 - \alpha)$       & $(h ,0)$   \\ \cline{2-4} 
                           & $(a, h+1, relevant)$      & $(1 - \gamma) \dot (1 - \alpha)$ & $(0 ,0)$   \\ \hline
\end{tabular}%
}
\caption{Transition Probability (Sapirshtein, Sompolinsky, \& Zohar, 2017)}
\label{tab:6}
\end{table}

Wang, Liew, \& Zhang (2019) plan to apply the off-policy based Q-learning to solve the problem, unfortunately the reason that Q-learning is used there does not mention in the original paper. Since Q-learning can only solve a linear reward function, the authors propose a new RL multi-dimensional algorithm based on the off-policy Q-learning. The new algorithm considers two Q-functions, i.e., a pair of $(Q^{(a)}(s,a), Q^{(h)}(s, a))$. At each time step, the adversary agent observes $(s_{t+1}, r_{t+1}^a, r_{t+1}^h)$ from the environment. Then two Q-functions are updated as follows:
\begin{align}\label{eq:6.3}
q^{(a)}(s_t, a_t) \leftarrow (1-\beta)q^{(a)}(s_t, a_t)+\beta[(r_{t+1}^{(a)}+\lambda q^{(a)}(s_{t+1}, a')]
\\
q^{(h)}(s_t, a_t) \leftarrow (1-\beta)q^{(h)}(s_t, a_t)+\beta[(r_{t+1}^{(h)}+\lambda q^{(h)}(s_{t+1}, a')]
\end{align}
\newline where $\beta \in (0, 1)$ is the learning rate, $\lambda$ is a number close to 1, $a' = arg max_{a}f(s_{t+1}, a)$. The current best action is chosen by the $\epsilon$ greedy strategy to maximize the objective function~\ref{eq:6.2} with the probability $1 - \epsilon$. A random action is chosen with the probability $\epsilon$. The random selection is involved to avoid trapping at local maximums. The parameter $\epsilon$ is determined by $\epsilon(s_t) = exp(-\frac{V(s_t)}{T_{\epsilon}})$, where $V(s_t)$ is the number of times that the state was visited and $T_{\epsilon}$ controls the speed of reducing $\epsilon$. 

Sensitivity analysis is applied to evaluate the designed optimal strategy and the simulation result is shown in Figure 4. After setting the discounted factor as 1, the paper concludes that the optimal mining strategy outperforms current mining strategies presented in Eyal \& Sirer (2014) and Sapirshtein, Sompolinsky, \& Zohar (2017).

\begin{figure}
    \centering
    \includegraphics[scale=0.57]{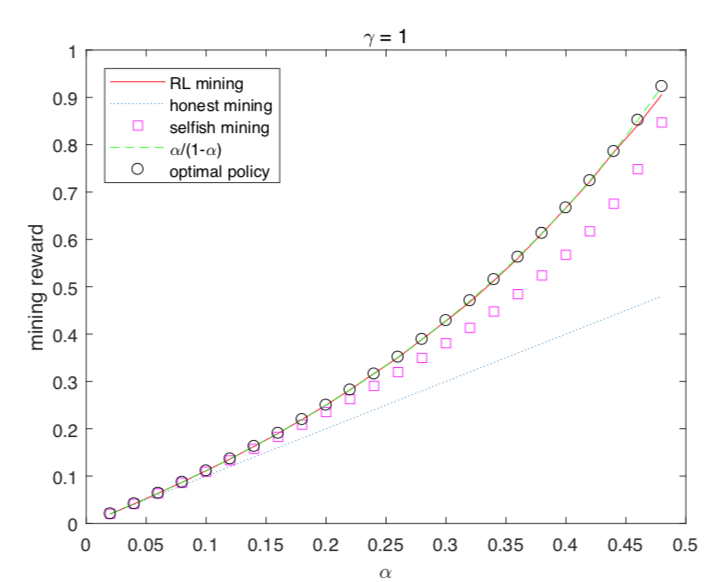}
    \caption{Simulation Result for Different Mining Strategies (Wang, Liew, \& Zhang, 2019)}
\end{figure}

\section{CONCLUSION AND FUTURE CHALLENGES}\label{sec:7}

The research we review either applies blockchain in a database to improve users' privacy in learning process; or uses machine learning to optimize  computer resource allocation or cryptocurrency investment decisions. The majority can be categorized as applying one technique to another; few is the actual integration of the two technologies. Hence, it is fair to say the current research is still very preliminary from an interdisciplinary perspective.

However, we expect new research lines to emerge in the following areas:
\begin{itemize}
\item  Design ``smart agents" with learning abilities to regulate the blockchain and detect abnormal behaviors. The former is especially important for consortium chain and private chain that require coordination among users, while the latter is critical for public chain;
\item The learning-based analysis of blockchain-based system is rare. From financial systems to supply chains, there is an enormous amount of data available to evaluate the performance of the decentralized structure of blockchain compared with the traditional centralized one. Learning-based analysis can shed insights on the mechanism design of the blockchain structures and provide on-time forecasting models; 
\item Blockchain to allow anonymously data sharing. With the development of IOT and wearable device, the privacy issue catches more and more attention of users. Combining with data fusion, we can design multiple-layer blockchain structures that allow sophisticated authorization of data for different users.
\item The blockchain mining activity could be considered as an MDP process. Although there exist a few works related to finding the optimal mining strategy using single-agent reinforcement learning, individual mining is not as popular as pool mining in reality. Specifically, miners collaborate and compete with each other to mine blocks. A multi-agent reinforcement learning (MARL) with a mixed setting of collaborative and competitive agents is more suitable to model the complex pool mining activity and helps miners find the optimal mining strategies in the future.
\item Cryptocurrency plays an important role, especially in the public chain. Different chains have their unique cryptocurrency. Now cryptocurrency or cryptocurrency portfolio is an investment option similar to other financial products. Some works have studied cryptocurrency price prediction using supervised learning techniques, but only a few of them explore potentials of RL or deep RL. In many cases, RL and deep RL have a better in financial forecasts, e.g., stock price prediction, since historical data cannot reflect the current market, which further results in poor prediction performance of future price changes. We expect that more works adopting RL, deep RL, or inverse RL to study the investment return of cryptocurrencies emerge soon.

\end{itemize}

\end{document}